\global\let\ifmypprint\iffalse 
\def\mypprint{\global\let\ifmypprint\iftrue}
\global\let\iftorefs\iffalse
\def\torefs{\global\let\iftorefs\iftrue}
\global\let\dofloatfig\iffalse
\def\floatthefig{\let\dofloatfig\iftrue}
\ifmypprint
\documentstyle[twocolumn,prb,aps]{revtex}
\floatthefig
\else
  \documentstyle[preprint,prb,aps]{revtex}
  \iftorefs 
    \catcode`\@=11 
    
    \def\figure{\let\@capwidth\columnwidth\@float{figure}}
    \let\endfigure\end@float
    \@namedef{figure*}{\let\@capwidth\textwidth\@dblfloat{figure}}
    \@namedef{endfigure*}{\end@dblfloat}
    \catcode`\@=12 
     \floatthefig 
  \fi
\fi

\def\unvz{{\bf \hat {z}}}	% z unit vector
\def\lapl{\nabla^2}		% Laplacian
\def\grad{\nabla}		% Gradient
\def\pd#1,#2;{\frac{\partial {#1}}{\partial {#2}}}
\def\d#1,#2;{\frac{d {#1}}{d {#2}}}
\def\dt{\frac{d}{dt}}
\def\px{\partial_x}
\def\py{\partial_y}
\def\sf{{{\raise.5ex\hbox{$\chi$}}}} % Streamfunction
\def\Td{T}			% Temperature deviation from conduction prof.
\def\thermc{\kappa}		% Thermal conductivity
\def\visc{\nu}			% Viscosity
\def\Ra{R}			% Rayleigh number
\def\Pr{\sigma}			% Prandtl number
\def\Nu{{\rm Nu}}		% Nusselt number
\def\Ekin{{K}}			% Kinetic energy of fluid
\def\Epot{{U}}			% Potential energy
\def\Etot{{E}}			% Total energy
\def\l{\left}
\def\r{\right}

\def\dsum#1,#2;{\sum_{\scriptscriptstyle#1\atop\scriptscriptstyle#2}}
\def\PB#1,#2;{\l[ #1 \mathchar"213B #2 \r]}
\def\avg#1;{{\l\langle #1 \r\rangle}}
\def\xavg#1;{{\overline {#1}}}
\def\ar{{L}}
\def\iar{{\alpha}}
\def\sf{{\psi}}
\def\modsf{{A_\sf}}
\def\modTd{{A_\Td}}
\def\dsum#1,#2;{\sum_{\scriptscriptstyle#1\atop\scriptscriptstyle#2}}
\def\qcd{{q_{\rm cd}}}	% Heat transported by conduction
\def\qcv{{q_{\rm cv}}}	% Heat transported by convection

\input epsf

\begin{document}

\draft

\bibliographystyle{pfa}

\title{Energy-Conserving Truncations for Convection with Shear Flow}

\author{Jean-Luc Thiffeault\thanks{e-mail: jeanluc@hagar.ph.utexas.edu} 
	and Wendell Horton\thanks{e-mail: horton@hagar.ph.utexas.edu} }

\address{Institute for Fusion Studies and University of Texas, Austin, 
	Texas, 78712-1060}

\date{\today}

\maketitle

\widetext

\begin{abstract}

A method is presented for making finite Fourier mode truncations of the
Rayleigh--B\'enard convection system that preserve invariants of the full
partial differential equations in the dissipationless limit.  These
truncations are shown to have no unbounded solutions and provide a description
of the thermal flux that has the correct limiting behavior in a steady-state.
A particular low-order truncation (containing~7 modes) is selected and
compared with the~6 mode truncation of Howard and
Krishnamurti,\cite{Howard1986} which does not conserve the total energy in the
dissipationless limit.  A numerical example is presented to compare the two
truncations and study the effect of shear flow on thermal transport.

\end{abstract}


\narrowtext
%\twocolumn

\section{Introduction}
\label{sec:intro}

In a horizontal layer of fluid with fixed higher temperature on the bottom
boundary and fixed lower temperature on the top boundary, cellular convective
flow occurs for a certain range of Rayleigh number~$\Ra$ and Prandtl
number~$\Pr$.\cite{Chandra} Such states of thermal convection are ubiquitous
in nature, occuring in slightly modified form in the atmosphere, the ocean,
the earth's mantle, and in the convection zone of the interior of stars.  It
was originally believed that flows in a finite container should scale as its
vertical dimension.  However, experiments have shown that thermal convection
in a horizontal layer of fluid heated from below can show motions spanning the
largest horizontal dimension of the container,\cite{Krishnamurti1981} known as
{\it shear flows}.  Since the experimental setting precluded any externally
imposed shear forces, it is concluded that these shear flows are driven by a
Reynolds stress tensor with non-vanishing horizontal average.  This behavior
is also seen in numerical experiments.\cite{Prat1995}

In this paper we make truncations of the Rayleigh--B\'enard system using the
{\it Galerkin} (or {\it spectral}) method.\cite{Orszag1971} The chosen basis
is the standard Fourier one, because
of its great simplicity and the fact that it is especially well-suited to the
stress-free boundary conditions. One of the major points to be addressed is how
to make a truncation that retains the truncated invariants of the full Partial
Differential Equations (PDE's) in the dissipationless limit, and whether or
not this has any effect on simulations in the dissipative case.  We will show
that energy-conserving truncations have several advantages over their
non-energy-conserving counterpart.
\ifmypprint\pagebreak \vspace*{1.232in} \noindent \fi

Section~\ref{sec:model} is devoted to a presentation of the equations
governing the system and the preserved quantities of the equations of motion
in the dissipationless limit are displayed.  In Section~\ref{sec:modexpansion}
we use the Galerkin method to expand the stream function and temperature field
into a complete set of modes, deriving a set of coupled Ordinary Differential
Equations (ODE's).  The condition under which a mode truncation preserves the
invariants of the dissipationless PDE's is obtained.
Section~\ref{sec:7odemodel} presents a low-order model for shear flow
generation, which is an extension of a model presented in
Ref.~\onlinecite{Howard1986}.  The model is compared with its predecessor and
the advantages are explicitly shown.  Finally, Section~\ref{sec:conclusion} is
a summary of the main arguments of the paper.

\section{Model Equations}
\label{sec:model}

The two-dimensional Rayleigh--B\'enard problem for an incompressible fluid
is governed in the Boussinesq approximation by
\begin{mathletters}
	\label{eq:Bous}
\begin{eqnarray}
	\pd\lapl\sf,t; + \PB\sf,\lapl\sf; &=&
		\pd\Td,x; + \visc \grad^4 \sf, 
	\label{eq:Boussf} \\
	\pd\Td,t; + \PB\sf,\Td; &=& \pd\sf,x; + \thermc \lapl \Td ,
	\label{eq:BousT}
\end{eqnarray}
\end{mathletters}
\noindent
where $\sf$ is the stream function, $\Td$ is the deviation of the temperature
from a linear conduction profile, $\visc$ is the kinematic viscosity, and
$\thermc$ is the thermal conductivity.  All the quantities are dimensionless,
the Prandtl number~$\Pr$ is~$\visc/\thermc$, and the Rayleigh number~$\Ra$
is~\hbox{$1/\visc \thermc$}.  The horizontal coordinate is~$x$ and the
vertical one is~$y$, with~\hbox{$(x,y) \in [0,2\pi \ar]\times[0,\pi]$}.  (The
choice~\hbox{$y \in [0,\pi]$}, as opposed to~\hbox{$y \in [0,1]$}, leads to a
Rayleigh number smaller by a factor of~$\pi^4$.)  The Poisson bracket used is
defined as~\hbox{$\PB A,B; = \px A\,\py B - \py A\, \px B$}, and the velocity
field is given in terms of the stream function by~\hbox{${\bf v} = \unvz
\times \grad\sf = (-\py\sf,\px\sf)$}.  We assume the fluid is periodic in~$x$
and has stress-free boundary conditions at the top and bottom walls:
\begin{equation}
	\sf = \lapl\sf = \px\sf = \Td = 0,\ \ \ 
		{\rm for}\ y = 0 \ {\rm or} \ \pi.
	\label{eq:sfBC}
\end{equation}

In the dissipationless limit,~\hbox{$\visc = \thermc = 0$},
Eqs.~(\ref{eq:Bous}) admit an infinite number of conserved quantities (this is
a general feature of noncanonical infinite-dimensional Hamiltonian
systems---for a discussion of the invariants of an analogous set of equations,
see Ref.~\onlinecite{Morrison1984}).  Here we shall concern ourselves with the
total energy,~$\Etot$, given by
\begin{eqnarray}
	\Etot = \frac{1}{2} \avg (\grad \sf)^2; - \avg y\, \Td; =
	\Ekin + \Epot,
	\label{eq:Edef}
\end{eqnarray}
where~$\Ekin$ is the kinetic energy,~$\Epot$ is the potential energy, and the
angle brackets denote the integral over the fluid domain.  In the
dissipationless limit, the time derivative of~$\Etot$ is
\begin{eqnarray}
	\dt\Etot &=& -\avg \sf{\pd\lapl\sf,t;}; - \avg y\,{\pd\Td,t;}; 
		\nonumber \\
	&=& \avg \sf \l({\PB \sf,\lapl\sf;} - {\px\Td}\r); 
		+ \avg y \l({\PB \sf,\Td;} - {\px\sf}\r); \nonumber \\
	&=& -\avg \sf\, {\px\Td}; + \avg \sf\, {\PB \Td,y;}; = 0\, ,
	\label{eq:expecons}
\end{eqnarray}
\noindent
showing that the total energy is conserved (we have set surface terms to zero
in Eq.~\ref{eq:expecons}).  The total internal (thermal) energy of the
fluid,~$\avg \Td;$, is also conserved.

\section{Mode Expansion}
\label{sec:modexpansion}

In this Section we look at truncations of system~(\ref{eq:Bous}) and their
properties.  In the first part we derive the equations of motion for modes of
the truncations (Section~\ref{subsec:ODEs}).  In Section~\ref{subsec:presinv}
we examine the behavior of the invariants mentioned in
Section~\ref{sec:model} after they are truncated and show (in a manner similar
to Ref.~\onlinecite{Treve1982}) that they can be made to remain invariant by
adding certain modes to the system.  Such truncations will be called~{\it
energy-conserving}.  Finally, in Section~\ref{subsec:bound} we examine the
important properties of these truncations, namely that the truncated system
has no singular solutions and that the thermal flux is properly modeled.

\subsection{Derivation of the ODE's}
\label{subsec:ODEs}

To turn the system of partial differential equations~(\ref{eq:Bous}) into
ordinary differential equations, we use the following normal mode expansions
for the $\sf$ and~$\Td$ fields:
\begin{mathletters}
\label{eq:fieldexp}
\begin{eqnarray}
	\sf(x,y,t) &=& \sum_{(m,n) \in \modsf} \,\sf_{mn}(t)\, 
		e^{i (m \iar x + n y)},
		\label{eq:sfexp} \\
	\Td(x,y,t) &=& \sum_{(m,n) \in \modTd} \,\Td_{mn}(t)\, 
		e^{i (m \iar x + n y)}
		\label{eq:Texp},
\end{eqnarray}
\end{mathletters}
\noindent
where $\iar \equiv 1/\ar$ is the inverse aspect ratio.  The summations are
over some sets~$\modsf$ and~$\modTd$ of modes (i.e., ($m,n$) pairs, where
both~$m$ and~$n$ can be negative or zero).  If both of these sets are infinite
and contain all possible~\hbox{($m,n$)} pairs, then the equalities hold in
(\ref{eq:fieldexp}); otherwise, the expansion is a truncation.
Expansion~(\ref{eq:fieldexp}) is more general than those used in
Refs.~\onlinecite{Curry1978,Treve1982,Curry1984} in two ways: first, it allows
for a variable phase in the rolls (by allowing the~$\sf_{mn}$'s to be complex)
and second, the expansion admits a non-vanishing shear flow part
(the~$\sf_{0n}$ modes).

The reality of the fields and the stress-free boundary 
conditions~(\ref{eq:sfBC}) lead to
\begin{mathletters}
\label{eq:modeBC}
\begin{eqnarray}
	\sf_{mn} = \sf_{-m,-n}^* = -\sf_{m,-n}\, ,\\
	\Td_{mn} = \Td_{-m,-n}^* = -\Td_{m,-n}\, ,
\end{eqnarray}
\end{mathletters}
\noindent
so that if, say,~$\sf_{11}$ is in~$\modsf$, then so
are~$\sf_{1,-1}$,~$\sf_{-1,1}$, and~$\sf_{-1,-1}$, and similarly for~$\modTd$.
Note that~$\modsf$ and~$\modTd$ need not contain the same modes.  For
convenience we define~\hbox{$\rho_{mn} \equiv \iar^2 m^2 + n^2$} to be the
eigenvalues of the operator~$-\lapl$.  If we insert Eqs.~(\ref{eq:fieldexp})
into the Boussinesq equations~(\ref{eq:Bous}), we obtain the following set of
coupled nonlinear ODE's:
\begin{mathletters}
\label{eq:ODEs}
\begin{eqnarray}
	\dt\sf_{mn} &=& -\visc\rho_{mn}\sf_{mn}
		- i \frac{\iar \, m}{\rho_{mn}} \Td_{mn} \label{eq:sfmnevol} \\
		&& \mbox{} + \!\!\!\dsum{m'+m''=m},{n'+n''=n};
		\!\!\!\iar\, {\scriptstyle (m'n''-m''n')}
		\frac{\rho_{m''n''}}{\rho_{mn}} 
		\,\sf_{m'n'} \,\sf_{m''n''}\ ,
		\nonumber \\
	\dt\Td_{mn} &=& -\thermc\rho_{mn}\Td_{mn} 
		+ i \iar \, m \,\sf_{mn}  \label{eq:Tmnevol} \\
		&&\mbox{} + \!\!\!\dsum{m'+m''=m},{n'+n''=n};
		\!\!\!\iar\, {\scriptstyle (m'n''-m''n')}
		\,\sf_{m'n'} \,\Td_{m''n''}\ . \nonumber
\end{eqnarray}
\end{mathletters}

\subsection{Preservation of the Invariants}
\label{subsec:presinv}

The kinetic and potential energies of Eq.~(\ref{eq:Edef}) have the expansions
\begin{mathletters}
\label{eq:Eexp}
\begin{eqnarray}
	\Ekin &=& \case{1}{2} \sum_{m,n} \rho_{mn} \l| \sf_{mn} \r|^2\, ,
		\label{eq:Ekinexp} \\
	\Epot &=& i \sum_{p \ne 0} \frac{(-1)^{p}}{p} \,\Td_{0p}\ .
		\label{eq:Epotexp}
\end{eqnarray}
\end{mathletters}
\noindent
We now ask whether the total energy~$\Etot$ is still conserved in the
dissipationless limit for a truncated system.  Taking the time derivative of
Eqs.~(\ref{eq:Eexp}) and using Eqs.~(\ref{eq:ODEs}) with~\hbox{$\visc =
\thermc = 0$}, we obtain after some manipulation:
\begin{eqnarray}
	\dt\Etot &=& \dt(\Ekin + \Epot) \nonumber \\
	&=& i\iar\sum_{m,n}m\,\sf_{mn}\,\Td_{mn}^*
		+ i\iar\dsum{m,n},{p\ne 0};
		(-1)^p\,m\, \sf_{mn}\,\Td_{m,n-p}^*\, , \nonumber \\
	&=& i\iar\sum_{m,n,p} (-1)^p\,m\, \sf_{mn}\,\Td_{m,n-p}^*\, .
	\label{eq:truncEcons}
\end{eqnarray}
Let~\hbox{$N \equiv \{\max\, (n)\, |\, (m,n) \in \modsf \cup \modTd \}$}, i.e.,
the maximum vertical mode number included in the truncation.  If we assume the
sum over~$p$ runs from~$-N'$ to~$N'$, we can write Eq.~(\ref{eq:truncEcons})
as
\begin{equation}
	\dt\Etot = i\iar\sum_{m}\sum_{n=-N}^{N}\sum_{p=-N'}^{N'} 
		(-1)^p\,m\, \sf_{mn}\,\Td_{m,n-p}^*\, .
\end{equation}
Now replace~$p$ by~\hbox{$s=n-p$}:
\begin{equation}
	\dt\Etot = i\iar\sum_{m}\sum_{n=-N}^{N}\sum_{s=-N'+n}^{N'+n} 
		(-1)^{n-s}\,m\, \sf_{mn}\,\Td_{m,s}^*\, ,
\end{equation}
and note that the maximum lower bound for~$s$ is~\hbox{$-N'+N$}
when~\hbox{$n=N$}, while the minimum upper bound is~\hbox{$N'-N$}
when~\hbox{$n=-N$}.  If~$N'=2N$,~\hbox{$s \in [-N,N]$} always (since
for~\hbox{$|s| > N$} the mode is not included in the truncation and so is made
to vanish), and we can use the symmetries given by Eq.~(\ref{eq:modeBC}) to
show that~$d\Etot/dt$ vanishes.\cite{Treve1982} Hence, we must have~$p$
running from~$-2N$ to~$2N$, which from Eq.~(\ref{eq:Epotexp}) implies adding
the modes~\hbox{$\Td_{0,-2N} \ldots \Td_{0,2N}$} to~$\modTd$.

For the internal energy,~\hbox{$\avg\Td;$}, the expansion is
\begin{equation}
	\avg\Td; = \frac{2i}{\pi} \sum_{p\ {\rm odd}}\frac{\Td_{0p}}{p}\, ,
\end{equation}
and its time derivative in the dissipationless limit is
\begin{equation}
	\dt\avg\Td; = \frac{2\iar i}{\pi} \sum_{m,n}\sum_{p\ {\rm odd}}
		m\, \sf_{mn}\,\Td_{m,n-p}^*\, .
	\label{eq:dtTexp}
\end{equation}
Comparing Eq.~(\ref{eq:dtTexp}) with Eq.~(\ref{eq:truncEcons}), we see that
Eq.~(\ref{eq:dtTexp}) vanishes under the same condition as the total
energy~$\Etot$.\cite{Treve1982}

\subsection{Properties of the Truncations}
\label{subsec:bound}

To show that the truncated systems obtained in Section~\ref{subsec:presinv}
have bounded solutions for all times~$t>0$, we consider the
quantity~$Q$:\cite{Howard1986}
\begin{equation}
	Q \equiv \Ekin + 2\sum_{m,n>0} \l|\Td_{mn}\r|^2
		+ \sum_{n>0} \l(\Td_{0n}^i - \frac{2}{n}\r)^2,
	\label{eq:Qdef}
\end{equation}
where~$\Ekin$ is the kinetic energy defined previously.  The quantity~$Q$ is
non-negative and it includes all the modes in the truncation such that if any
of them diverges, then~$Q$ diverges.  Thus, if~$Q$ is bounded from above then
the truncated system has no unbounded solutions.  Taking the time derivative
of~$Q$, with the viscosity and thermal conductivity nonzero, we can write
\begin{equation}
	\dt Q \le -\min\{2\visc,\thermc\}Q + 4\thermc N_0\ ,
	\label{eq:Qbound}
\end{equation}
with $N_0$ being the number of $\Td_{0|n|}$ modes included in the truncation.
For \hbox{$Q > 4\thermc N_0/\min\{2\visc,\thermc\}$}, we have \hbox{$dQ/dt
<0$}, and so $Q$ is bounded.\cite{JLTthesis1995}

We define the horizontally averaged vertical thermal flux as~\hbox{$q(y) \equiv
\qcv(y) + \qcd(y)$}, where~\hbox{$\qcv(y) = \xavg v_y\Td;$} is the
convective thermal flux and~\hbox{$\qcd(y) = \thermc(1-\xavg \py\Td;)$} is the
\dofloatfig
\begin{figure}
	\epsfxsize=3.4 truein
	\centerline{\epsffile{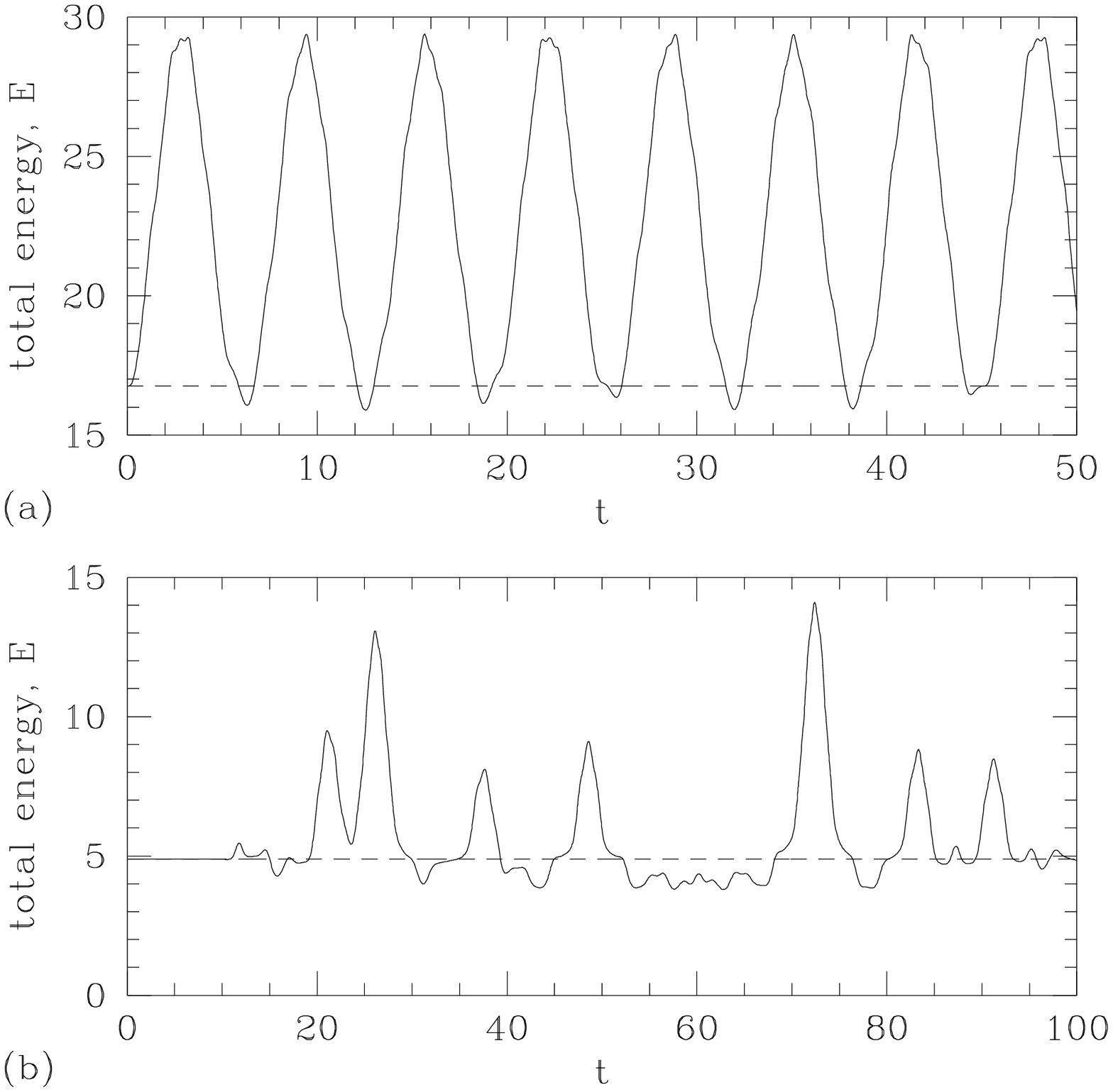}}
	%\smallskip
	\caption{Total energy~$\Etot$ in the dissipationless 
		limit (\hbox{$\visc = \thermc = 0$}) 
		for the~6-ODE (solid line) and~7-ODE (dashed
		line) models with~$\alpha=1.2$ and initial conditions 
		(a) $\sf_{11}^r = \sf_{01}^i = \sf_{12}^i = \Td_{11}^i = 1$;
		all other modes begin at~$10^{-5}$,
		(b) $\sf_{11}^r = \Td_{11}^i = 1$; all other modes begin 
		at~$10^{-8}$.}
	\label{fig:Econs}
\end{figure}
\noindent
\fi
conductive thermal flux (the overbar denotes an average over~$x$).  For
energy-conserving truncations one can write the expansion for~$q$
as~\cite{JLTthesis1995}
\begin{equation}
	q(y) = \avg q; - 2 \sum_{m > 0} \frac{\cos my}{m}\dt\Td_{0n}^i\, .
	\label{eq:efluxexp}
\end{equation}
In a steady-state situation this reduces to the expected result~\hbox{$q =
\avg q;$}, independent of~$y$, showing that the energy cannot ``pile up'' in
steady convection.  For a general truncation (for example, the truncations in
Refs.~\onlinecite{Howard1986,Curry1978,Curry1984}), one cannot write~$q$ in
the form given by Eq.~(\ref{eq:efluxexp}) and the thermal flux has an
unphysical~$y$ dependence in a steady-state.  General truncations can also
have unbounded solutions as is the case in Ref.~\onlinecite{Howard1986} for
large enough Rayleigh number.

\section{Low-order Truncations}
\label{sec:7odemodel}

A popular truncation of the Boussinesq equations (in the spirit of the Lorenz
model~\cite{Lorenz1963}) is the~6-ODE model given by Howard and
Krishnamurti~\cite{Howard1986} and used by other authors.\cite{Rucklidge1995}
It includes the~6 independent modes~$\sf_{01}^i$, $\sf_{11}^r$, $\sf_{12}^i$,
$\Td_{11}^i$, $\Td_{12}^r$, and $\Td_{02}^i$.  The Howard and Krishnamurti
truncation is the simplest one that allows for a nonzero shear flow
(the~$\sf_{01}^i$ mode).  It has a vanishing~$\avg T;$.  However, it is not
energy-conserving: it lacks the~$\Td_{04}^i$ mode.  We will add
this~$\Td_{04}^i$ mode to the~6-ODE model to obtain what we will call
the~7-ODE model.  Figure~\ref{fig:Econs} displays explicitly the
energy-conserving property of the~7-%
\dofloatfig
\begin{figure}
	\epsfxsize=3.4 truein
	\centerline{\epsffile{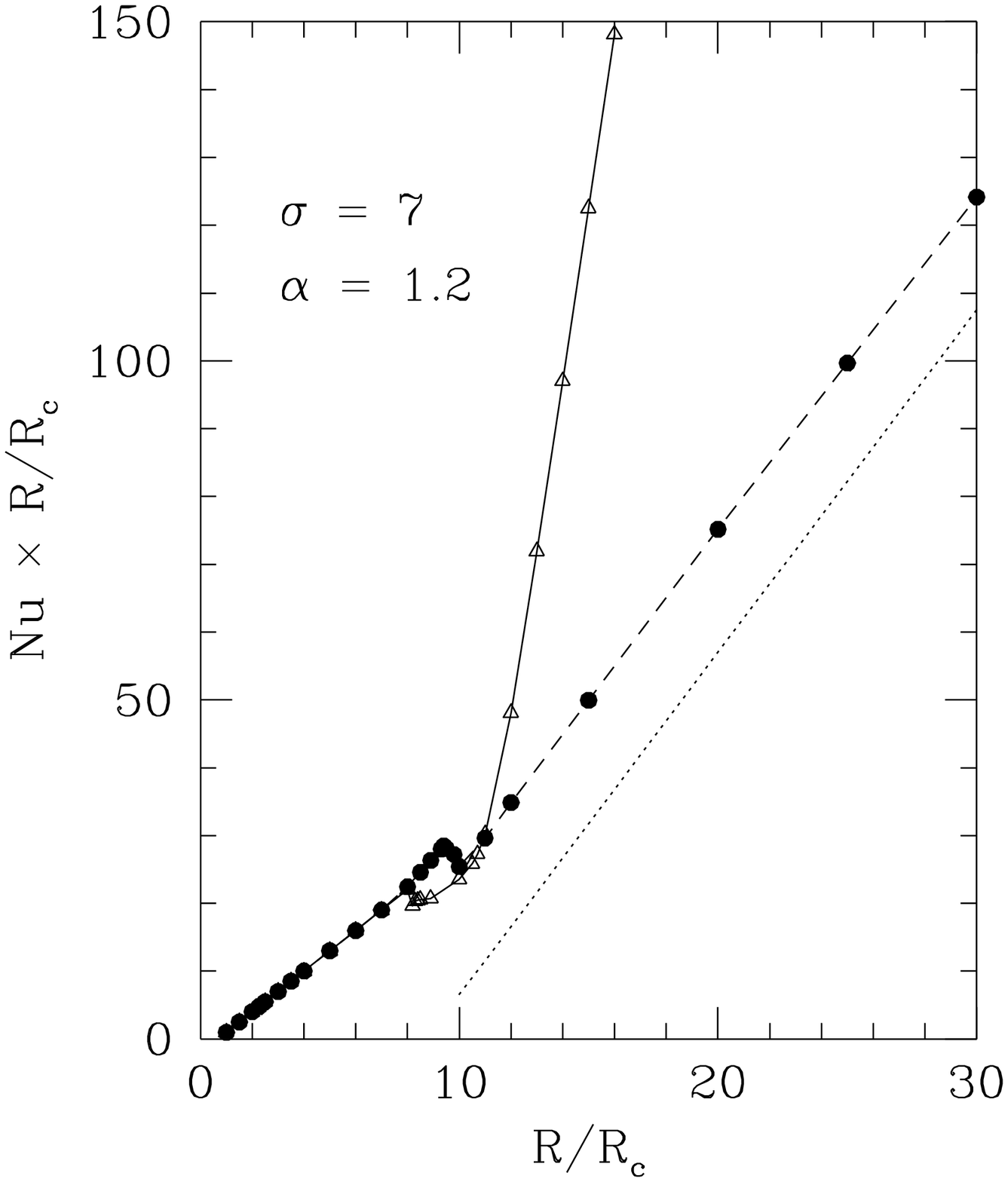}}
	%\smallskip
	\caption{Time-averaged~\hbox{$\Nu \times R/R_c$} evaluated at~$y=\pi$ 
		as a function of~$R/R_c$ for the~6-ODE (solid line, triangles)
		 and~7-ODE (dashed line, black dots) models.  The dotted line 
		has a slope of~$5.05$, corresponding to the experimental 
		results for~$\Pr=7$ in 
		Ref.~\protect\onlinecite{Krishnamurti1981}.}
	\label{fig:Comparison}
\end{figure}
\noindent
\fi
ODE model in the dissipationless limit.
Note that the nonconservation of energy for the~6-ODE model is not a small
effect:  Figure~\ref{fig:Econs} shows that the energy is not conserved by a
factor of~$1.8$ to as much as~$2.8$.

Figure~\ref{fig:Comparison} shows the time-averaged quantity~\hbox{$\Nu \times
R/R_c$} evaluated at the upper boundary plotted as a function of the Rayleigh
number scaled by~$R_c$, the critical Rayleigh number where the fluid at rest
becomes linearly unstable ($R_c = (1+\iar^2)^3/\iar^2 \simeq 10.088$ for
$\iar=1.2$).  The Nusselt number~$\Nu$ is the ratio of total heat transferred
to the heat conducted when the fluid is at rest, so that~$\Nu \times R/R_c$ is
a dimensionless measure of the thermal flux.\cite{Krishnamurti1981} The
thermal flux in the~6-ODE model (solid line) is seen to grow rapidly
after~\hbox{$R/R_c \simeq 11$}, whereas the same quantity for the~7-ODE model
has a slower growth (dashed line).  The dotted line has a slope of~$5.05$,
corresponding to the experimental result of Ref.~\onlinecite{Krishnamurti1981}
(the intercept is arbitrary) for~\hbox{$\Pr=7$} (the numerical results are
also for~\hbox{$\Pr=7$}).  The agreement between the experimental results and
that of the~7-ODE model is excellent.  However, since the experiment was done
with no-slip boundary conditions, caution should be taken in concluding the
accuracy of the model (the value of~$R_c$ used to scale the experimental curve
is the no-slip one).  The onset of shear flow (at~$R/R_c \simeq 8.1$) is
associated with a
decrease in thermal flux for the~6-ODE model but an increase of that quantity
for the~7-ODE model.  For smaller values of~$\Pr$, the two models behave in
a more similar fashion, as seen in Figure~\ref{fig:Comparison2} where the same
quantities are plotted for~\hbox{$\Pr=1$}.

Figure~\ref{fig:heatfluxHK} shows the thermal flux~$q$ for the~6-ODE model
\dofloatfig
\begin{figure}
	\epsfxsize=3.4 truein
	\centerline{\epsffile{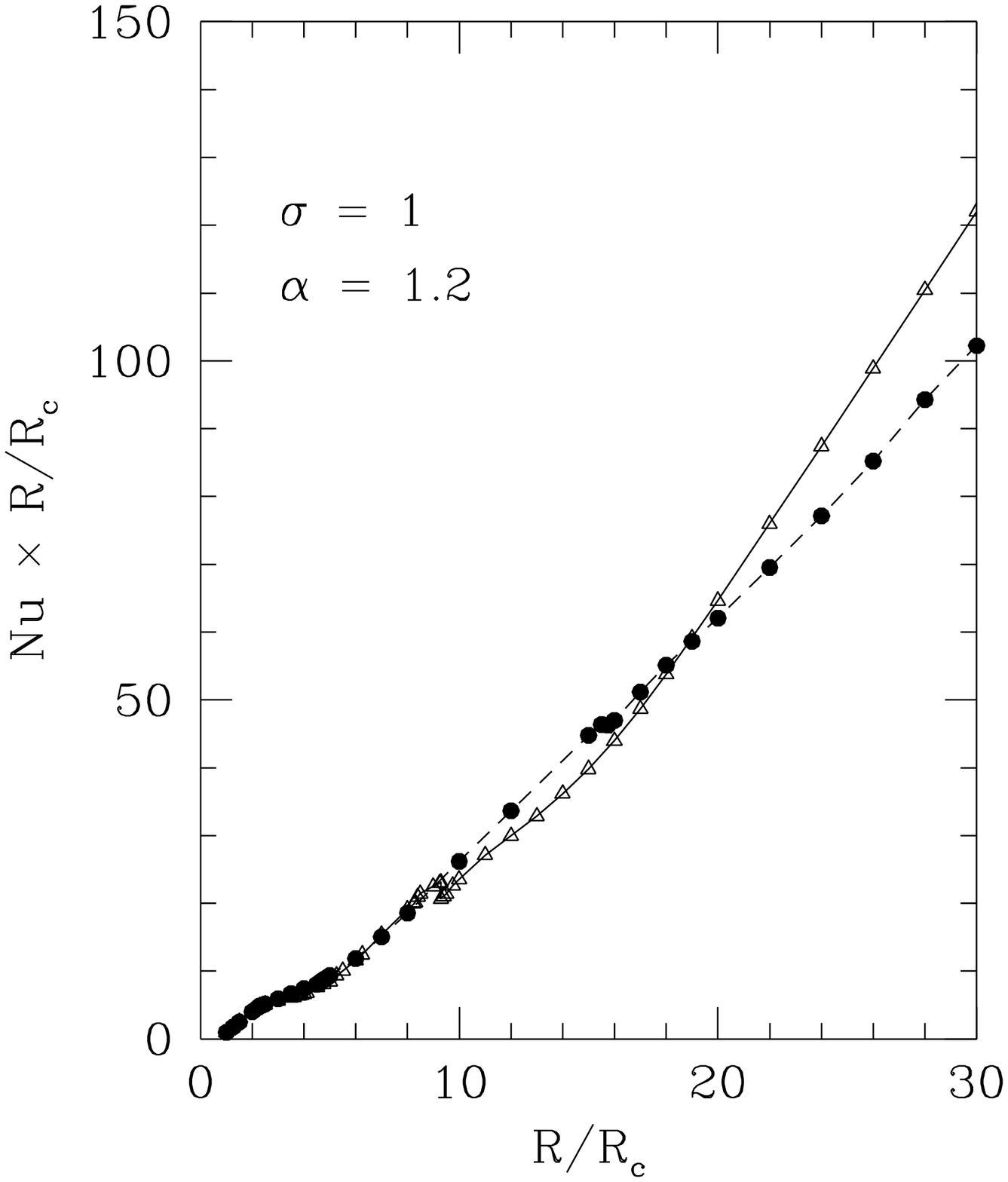}}
	%\smallskip
	\caption{Time-averaged~\hbox{$\Nu \times R/R_c$} evaluated at~$y=\pi$ 
		as a function of~$R/R_c$ for the~6-ODE (solid line, triangles)
		and~7-ODE (dashed line, black dots) models.}
	\label{fig:Comparison2}
\end{figure} 	
\noindent
\fi
evaluated at different values of~$y$ as a function of time ($R/R_c=3.5$,
$\Pr=1$, $\iar=1.2$).  The thermal flux after the system settles in a
steady-state is seen to depend on~$y$, which is unphysical since this would
lead to energy pile up (see Section~\ref{subsec:bound}).  However, as shown in
Figure~\ref{fig:heatfluxHKec} the thermal flux for the energy-conserving 7-ODE
model is independent of~$y$ when the system reaches a steady-state.

\dofloatfig
\begin{figure}
	\epsfxsize=3.4 truein
	\centerline{\epsffile{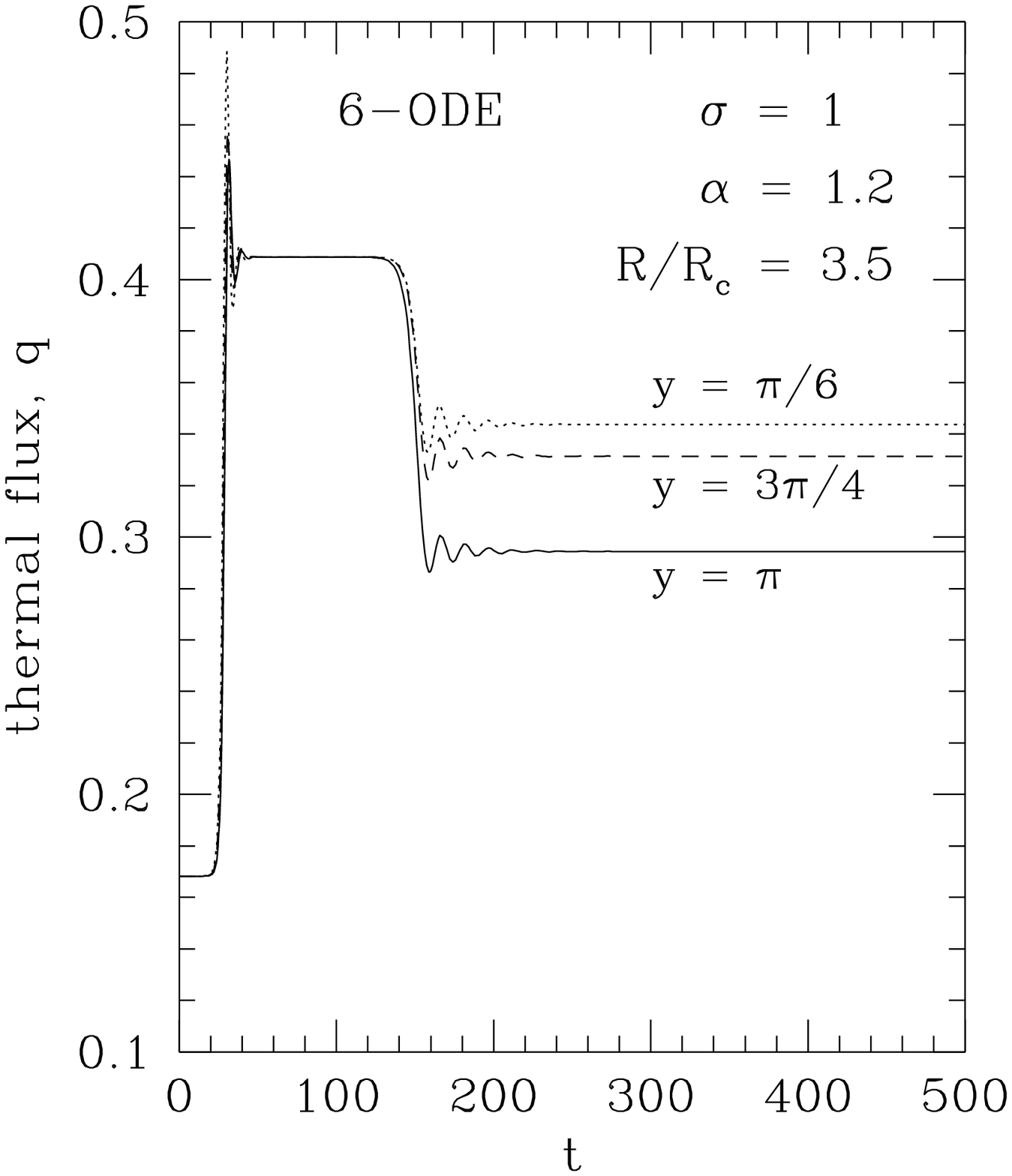}}
	\caption{Thermal flux~$q$ for the~6-ODE model evaluated at different 
	values of~$y$ as a function of time. The final, steady-state thermal 
	flux is seen to depend on~$y$.}
	\label{fig:heatfluxHK}
\end{figure}
\fi

\dofloatfig
\begin{figure}
	\epsfxsize=3.4 truein
	\centerline{\epsffile{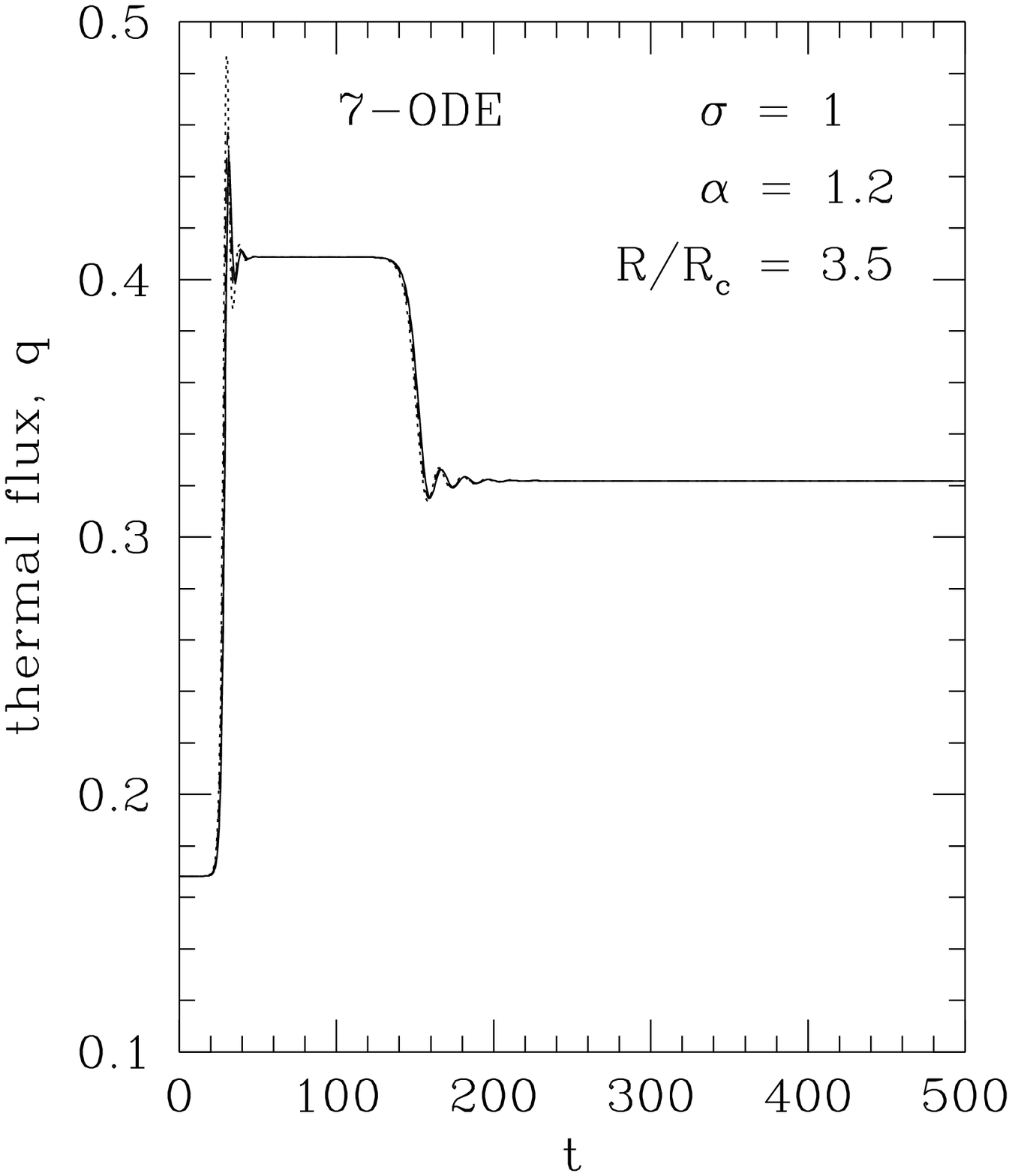}}
	\caption{Thermal flux~$q$ for the~7-ODE model evaluated at the same
	values of~$y$ as in Figure~\ref{fig:heatfluxHK} as a function of time.
	The final, steady-state thermal flux is independent of~$y$ as 
	expected physically.}
	\label{fig:heatfluxHKec}
\end{figure}
\fi

\section{Conclusions}
\label{sec:conclusion}

In this work we have developed a general method for generating
energy-conserving Galerkin approximations of the PDE's that describe
Rayleigh--B\'enard convection, using a generalization of the method of Treve
and Manley.\cite{Treve1982} The truncations allow for shear flow (zonal flows
independent of the horizontal coordinate,~$x$) and variable phase of the rolls
(breaking point symmetry with respect to the center of the rolls).  These
flows also have applications in tokamak plasmas, where it is thought that a
shear flow in the edge layer is responsible for the so-called
H-mode,\cite{Wagner1982} in which confinement is increased by a factor of two
over the normal, or L-mode phase.  Convection cells form as a result of the
nonlinear development of the Rayleigh--Taylor instability in regions of
unfavorable magnetic curvature.  Such convection cell turbulence is widely
observed in the edge of tokamak plasmas.\cite{Wooton1990} These vortices can
lead to the generation of a shear flow in a manner analogous to the
Rayleigh--B\'enard case\cite{Drake1992,Finn1992} and it is believed that this
flow creates a barrier to particle transport, thereby improving confinement.
The new truncations could help to provide a foundation for turbulence models
of L--H transitions such as in Ref.~\onlinecite{Sugama1995}.

There are essentially three arguments for using energy-conserving
approximations: First, the cascade of energy through the inertial range to the
dissipation scale is modeled without extraneous terms in the energy equations.
Thus, energy-conserving truncations are a good technique to reduce unphysical
numerical dissipation or sources and possible instabilities.  This makes the
truncations more closely related to the full equations, since in the full
PDE's the dissipation comes entirely from the linear terms and the nonlinear
parts of the equations conserve energy.  Note, however, that for the example
given (the~7-ODE model) there are too few modes to speak of an energy cascade
through an inertial range.  The second property is the correct description of
the thermal flux in the steady-state limit, even with dissipation.  This means
that the fact that a truncation does not conserve energy in the ideal limit
significantly affects the way the energy flows in the dissipative regime, and
so the energy-conserving truncations are relevant to the dissipative case.
Finally, the boundedness of solutions is a strong point since physically one
does not expect divergent behavior and hence the energy-conserving
truncations are more reliable.

In the dissipationless limit, we have demonstrated that energy conservation is
violated drastically for a typical truncation (the~6-ODE Howard and
Krishnamurti model\cite{Howard1986}), whereas it is conserved to machine
precision for the energy-conserving truncation (7-ODE model).  In the presence
of dissipation, the numerical results for the comparison of the two models for
Prandtl number~\hbox{$\Pr=7$} clearly show that the energy-conserving
truncation is much closer to experimental results.  If the system reaches a
steady-state, the~7-ODE model properly models the thermal flux, whereas
the~6-ODE model exhibits an unphysical dependence on the vertical coordinate.
Thus we believe that energy-conserving truncations represent the full system
more accurately.  Note, however, that the~7-ODE model is presented here as an
illustration of the energy-conserving truncation technique, as severely
truncated systems may yield numerical results far removed from the behavior of
the full system.\cite{Treve1982,Curry1984,Foias1983} For instance, in
Ref.~\onlinecite{Prat1995} the authors concluded that there are not enough
modes in the~6-ODE truncation to adequately model the bifurcations observed in
a full simulation of the flow.  Note also that Ref.~\onlinecite{Hermiz1995}
gives a different~7-ODE extension of the~6-ODE model that conserves total
vorticity in the dissipationless limit.  That model could be made
energy-conserving by adding to it a~$\Td_{06}^i$ mode.

\acknowledgments

The authors would like to thank Philip J. Morrison and Neil
J. Balmforth for useful discussions.  J.-L.T. acknowledges support
from the Natural Sciences and Engineering Research Council of Canada.

\dofloatfig\else
\begin{figure}
	\caption{Total energy~$\Etot$ in the dissipationless 
		limit~(\hbox{$\visc = \thermc = 0$}) 
		for the~6-ODE (solid line) and~7-ODE (dashed
		line) models with~$\alpha=1.2$ and initial conditions 
		(a) $\sf_{11}^r = \sf_{01}^i = \sf_{12}^i = \Td_{11}^i = 1$;
		all other modes begin at~$10^{-5}$,
		(b) $\sf_{11}^r = \Td_{11}^i = 1$; all other modes begin 
		at~$10^{-8}$.}
	\label{fig:Econs}
\end{figure}

\begin{figure}
	\caption{Time-averaged~\hbox{$\Nu \times R/R_c$} evaluated at~$y=\pi$ 
		as a function of~$R/R_c$ for the~6-ODE (solid line, triangles)
		 and~7-ODE (dashed line, black dots) models.  The dotted line 
		has a slope of~$5.05$, corresponding to the experimental 
		results for~$\Pr=7$ in 
		Ref.~\protect\onlinecite{Krishnamurti1981}.}
	\label{fig:Comparison}
\end{figure}

\begin{figure}
	\caption{Time-averaged~\hbox{$\Nu \times R/R_c$} evaluated at~$y=\pi$ 
		as a function of~$R/R_c$ for the~6-ODE (solid line, triangles)
		and~7-ODE (dashed line, black dots) models.}
	\label{fig:Comparison2}
\end{figure}

\begin{figure}
	\caption{Thermal flux~$q$ for the~6-ODE model evaluated at different 
	values of~$y$ as a function of time. The final, steady-state thermal 
	flux is seen to depend on~$y$.}
	\label{fig:heatfluxHK}
\end{figure}

\begin{figure}
	\caption{Thermal flux~$q$ for the~7-ODE model evaluated at the same
	values of~$y$ as in Figure~\ref{fig:heatfluxHK} as a function of time.
	The final, steady-state thermal flux is independent of~$y$ as 
	expected physically.}
	\label{fig:heatfluxHKec}
\end{figure}
\fi

\ifmypprint\else
	\vfill\eject
	\epsfxsize=6 truein  \centerline{\epsffile{Econsfig.ps}}
 	\vfill{\small Thiffeault and Horton, Fig.~1} \vfill\eject
	\epsfysize=6 truein  \centerline{\epsffile{Comparisonsig7.ps}}
	\vfill{\small Thiffeault and Horton, Fig.~2} \vfill\eject
	\epsfysize=6 truein  \centerline{\epsffile{Comparison2.ps}}
	\vfill{\small Thiffeault and Horton, Fig.~3} \vfill\eject
	\epsfysize=6 truein  \centerline{\epsffile{heatfluxHK.ps}}
	\vfill{\small Thiffeault and Horton, Fig.~4} \vfill\eject
	\epsfysize=6 truein  \centerline{\epsffile{heatfluxHKec.ps}}
	\vfill{\small Thiffeault and Horton, Fig.~5} \vfill\eject
\fi

\end{document}